\documentclass[aps,prd,reprint,amsmath,amssymb,superscriptaddress,nofootinbib,nolongbibliography,floatfix]{revtex4-2}
\usepackage{graphicx}
\usepackage{dcolumn}
\usepackage[utf8]{inputenc}
\usepackage{xcolor}
\usepackage[unicode=true,colorlinks,linkcolor=blue,anchorcolor=blue,urlcolor=blue,citecolor=blue]{hyperref}

\newcommand{\ii}{\mathrm{i}}

\newcommand{\pararrow}{\mathord{\buildrel{\lower3pt\hbox{$\scriptscriptstyle\leftrightarrow$}}\over {\partial}}} 
\newcommand{\pararrowk}[1]{\mathord{\buildrel{\lower3pt\hbox{$\scriptscriptstyle\leftrightarrow$}}\over {\partial}\hspace*{-0.18em}{}^#1}\hspace*{-0.18em} } 

\usepackage{physics}

\newcommand{\yd}{\Upsilon_1(3\,{}^3D_1)}

\newcommand{\qfnu}{\affiliation{College of Physics and Engineering, Qufu Normal University, Qufu 273165, China}}

\newcommand{\itp}{\affiliation{CAS Key Laboratory of Theoretical Physics, Institute of Theoretical Physics,\\ Chinese Academy of Sciences, Beijing 100190, China}}

\newcommand{\imp}{\affiliation{Institute of Modern Physics, Chinese Academy of Sciences, Lanzhou 730000, China}}

\newcommand{\snst}{\affiliation{School of Nuclear Science and Technology, University of Chinese Academy of Sciences, Beijing 101408, China}}

\newcommand{\scnt}{\affiliation{Southern Center for Nuclear-Science Theory (SCNT), Institute of Modern Physics, Chinese Academy of Sciences, Huizhou 516000, Guangdong,
China}}

\begin{document}
	
\title{Hidden-bottom hadronic transitions of $\Upsilon(10753)$}

\author{Shi-Dong Liu}\email{liusd@qfnu.edu.cn}\qfnu
\author{Zu-Xin Cai}\qfnu \imp
\author{Zhao-Sai Jia}\qfnu \itp
\author{Gang Li}\email{gli@qfnu.edu.cn} \qfnu \itp
\author{Ju-Jun Xie} \email{xiejujun@impcas.ac.cn} \imp \snst \scnt

\begin{abstract}

Assuming that the $\Upsilon(10753)$ is a $4S$-$3D$ mixed state, we investigated the hidden-bottom hadronic decays of the $\Upsilon(10753) \to \eta_b(1S)\omega(\eta^{(\prime)})$ via the intermediate meson loops. In a commonly accepted range of the model parameter $\alpha$ in the form factor, the predicted branching ratios may reach to the order of $10^{-3}$--$10^{-2}$. The relative ratio of the partial decay widths of the $\Upsilon(10753)\to\eta_b\eta^{(\prime)}$ to $\Upsilon(10753)\to \eta_b\omega$ is found to be dependent on the $\eta$-$\eta'$ mixing angle. In addition, we also calculated the ratios of the partial decays widths of the $\Upsilon(10753) \to \eta_b \omega$ to $\Upsilon(10753)\to \Upsilon(nS)\pi^+\pi^-$ ($n=1\,,2$), which are found to be around 0.4 and 0.2 for $n=1$ and $n=2$, respectively. These values are in accordance with the preliminary experimental results. The calculations presented here tend to favor the $\Upsilon(10753)$ as the $4S$-$3D$ mixture. We hope these predictions could be verified by the future BelleII experiments.

\end{abstract}

\date{\today}
	
	
	
	
	
	
\maketitle
	
\section{Introduction}\label{sec:intro}

Decays of the heavy quarkonia can provide us insight of the dynamics of quark interactions and the formation of hadrons~\cite{2005brambillaheavy,2008eichtenRMP80-1161,2011brambillaEPJC71-1534}. By means of comparing the theoretical predictions based on quantum chromodynamics (QCD) to the experimental data, we can test the validity of the model and improve our understanding of the strong interaction in the low energy region. Therefore, searching in experiments more new states in the $c\bar{c}$ and $b\bar{b}$ sectors and predicting their properties from different theoretical models are of particular importance. As summarized in the Review of Particle Data Group (PDG)~\cite{2022particledatagroupPoTaEP2022-083C01}, huge data samples have been accumulated. Among these samples, there are quite a few complex structures that cannot be interpreted by the conventional quark model, which are usually referred to as exotic or \textit{XYZ} states, e.g., the celebrated $X(3872)$  \cite{2003bellecollaborationPRL91-262001} and $Z_b(10610)/Z_b(10650)$ \cite{2012bellecollaborationPRL108-122001} (For review, see Ref.~\cite{2020PRbrambilla,2018guoRMP90-015004,2017lebedPiPaNP93-143,2019kalashnikovaP62-568,2016chenPR639-1,2023mengPR1019-1,2017baruJHEP2017-158}).

In the $b\bar{b}$ sector, there are only four documented bottomonium(-like) vector structures above the $B\bar{B}$ threshold: $\Upsilon(4S)$, $\Upsilon(10753)$, $\Upsilon(10860)$, and $\Upsilon(11020)$ in ascending order of their mass \cite{2022particledatagroupPoTaEP2022-083C01}. The $\Upsilon(10753)$ was newly observed in 2019 when the Belle collaboration updated measurement of the energy dependence of the $e^+e^-\to \Upsilon(nS)\pi^+\pi^-~(n=1\,,2\,,3)$  cross sections; the Breit-Wigner mass was measured to be $(10752.7\pm 5.9^{+0.7}_{-1.1})~\mathrm{MeV}$ with a width of $(35.5^{+17.6}_{-11.3}{}^{+3.9}_{-3.3})~\mathrm{MeV} $; its global significance was reported to be $5.2\sigma$ \cite{2019mizukJHEP2019-220}. The quantum numbers of the $\Upsilon(10753)$ are determined to be $J^{PC} = 1^{--}$ in view of the production processes. Three years later, this new structure $\Upsilon(10753)$ was also found in the $e^+e^-\to\omega\chi_{b1,b2}(1P)$ reactions at center-of-mass energy between 10.653 GeV and 10.805 GeV~\cite{2023belleiicollaborationPRL130-091902}.

The $\Upsilon(10753)$ is particularly peculiar since its mass does not fit with any possible conventional bottomonia predicted by various theoretical models~\cite{2020liEPJC80-59,2010badalianPAN73-138,2015godfreyPRD92-054034,2016segoviaPRD93-074027,2018wangEPJC78-915}. Precisely speaking, its mass is located between the theoretically predicted masses of the $\Upsilon(5S)$ and $\yd$ \cite{2010badalianPAN73-138,2015godfreyPRD92-054034,2016segoviaPRD93-074027,2018wangEPJC78-915,2020liEPJC80-59}: 
\begin{align*}
	m_{\Upsilon(10753)}-m_{\yd} \gtrsim 50~\mathrm{MeV}\,,\\ m_{\Upsilon(5S)}-m_{\Upsilon(10753)}\gtrsim 50~\mathrm{MeV}\,.
\end{align*}
Due to the rather small dielectron widths of $\sim$ $1~\mathrm{eV}$ for pure $n{}^3D_1$ states~\cite{2009badalianPRD79-037505,2010badalianPAN73-138}, a direct observation of a pure $D$-wave vector state in $e^+e^-$ experiments is impossible. Evidently, assigning the $\Upsilon(10753)$ as the pure $\yd$ is difficult. The alternative assignment of the $\Upsilon(10753)$ as a $\Upsilon(5S)$ state also makes conflict, since under such a assignment the $\Upsilon(10860)$ that possesses most of the properties of the $5S$-wave state would have no space in the bottomonium spectroscopy~\cite{2020liEPJC80-59}.

Previous theoretical efforts~\cite{2019wangCPC43-123102,2023bicudoPRD107-094515,2021bicudoPRD103-074507,2020aliPLB802-135217,2020PRbrambilla,2021tarruscastellaPRD104-034019,2022huskenPRD106-094013,2020liangPLB803-135340,2022kherEPJP137-357,2020gironPRD102-014036,2020chenPRD101-014020,2020liEPJC80-59,2021liPRD104-034036,2022liPRD105-114041,2022baiPRD105-074007} have been devoted to resolving the nature of the $\Upsilon(10753)$, including three main unusual interpretations: tetraquark state \cite{2019wangCPC43-123102,2020aliPLB802-135217,2021bicudoPRD103-074507,2023bicudoPRD107-094515}, hybrid bottomonium with excited gluonic degrees of freedom \cite{2020PRbrambilla,2021tarruscastellaPRD104-034019}, and $S$-$D$ mixture~\cite{2020liEPJC80-59,2021liPRD104-034036,2022liPRD105-114041,2022baiPRD105-074007,2010badalianPAN73-138}. Specifically, Wang \cite{2019wangCPC43-123102} took the $\Upsilon(10753)$ as a tetraquark state and predicted, using the QCD sum rules, its mass and width to be $(10.75\pm 0.10)~\mathrm{GeV}$ and $(33.60^{+16.64}_{-9.45})~\mathrm{MeV}$, which are in excellent agreement with the experimental results~\cite{2019mizukJHEP2019-220,2022particledatagroupPoTaEP2022-083C01}. Meanwhile, it was predicted that the open-bottom decays of the $\Upsilon(10753)$ are $B\bar{B}$ and $\bar{B}^*\bar{B}^*$ modes with branching fraction up to $92\%$, while the hidden-bottom decays are $\eta_b(1S)\omega$ and $\Upsilon(1S)\pi^+\pi^-$ modes with fraction of about $8\%$. Li et al \cite{2020liEPJC80-59} explained the $\Upsilon(10753)$ as the $5S$-$4D$ mixture with a mixing angle of (20--30)${}^\circ$ and obtained a different prediction that the open-bottom decay modes of the $B^*\bar{B}^*$ and $B\bar{B}^*+\mathrm{c.c.}$ are dominant while the mode $B\bar{B}$ is of insignificance. Furthermore, Liu Xiang's group \cite{2021liPRD104-034036,2022liPRD105-114041,2022baiPRD105-074007} suggested that the $\Upsilon(10753)$ is the $4S$-$3D$ mixed state and predicted the branching fractions of the hidden-bottom hadronic decays of $\Upsilon(10753) \to \Upsilon(1S)\eta^{(\prime)}$, $h_b(1P)\eta$, $\chi_{bJ}\omega$, $\Upsilon(1{}^3D_{J})\eta$, and $\Upsilon(nS)\pi^+\pi^-$ ($J=0\,,1\,,2\,;n=1\,,2\,,3$) via the intermediate meson loop mechanism.

A search in the literature yields that the investigations on the decays of the $\Upsilon(10753)$ are scarce so that it is necessary to study more possible decay modes to provide a theoretical basis for the future experiments. In this work, we study the processes of the $\Upsilon(10753)$ decaying into $\eta_b(1S) \omega$, $\eta_b(1S) \eta$, and $\eta_b(1S) \eta'$ considering the contributions of the intermediate meson loops, which have been widely used in the productions and decays of the bottomonium(-like) states \cite{2023wuPRD107-034028,2023wangEPJC83-186,2021liPRD104-034036,2022baiPRD105-074007,2022liPRD105-114041,2019wuPRD99-034022,2018zhangPRD97-014018,2018huangPRD97-094018,2018huangPRD98-054008,2017huangEPJC77-165,2017chenPRD95-034022,2016wangPRD94-094039,2016huoEPJC76-172,2015liPRD91-034020,2014liIJMPCS29-1460231}.

The rest of the paper is organized as follows. In Sec.~\ref{sec:formula},
we present the theoretical framework used in this work. Then in Sec.~\ref{sec:results} the numerical results are presented, and a brief summary is given in Sec.~\ref{sec:summary}.
	
\section{Theoretical Considerations}
\label{sec:formula}

\begin{figure*}[htbp]
	\centering
	\includegraphics[width=0.75\linewidth]{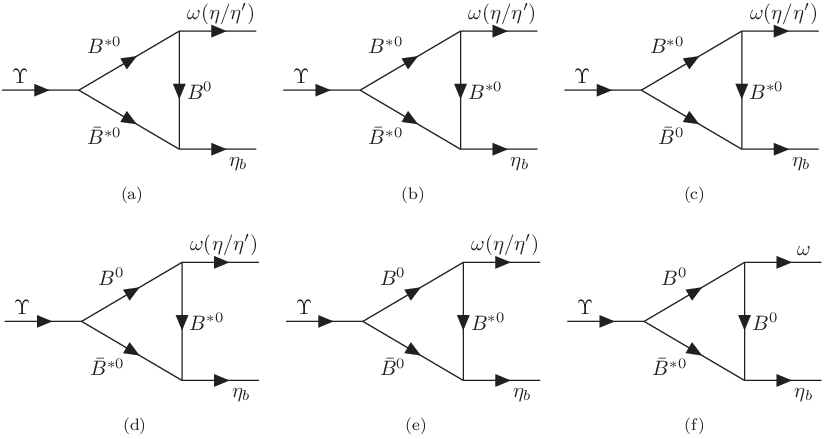}
	\caption{Feynman diagrams for the decay $\Upsilon(10753)\to \eta_b \omega(\eta^{(\prime)})$ in the neutral bottom meson loop mechanism. The corresponding charged and strange bottom meson loops are not shown but included in the calculations. The symbol $\Upsilon$ in the diagrams stands for the $\Upsilon(4S)$ and $\yd$.}
	\label{fig:feyndiags}
\end{figure*}

Figure \ref{fig:feyndiags} shows the neutral bottom meson loops devoted to the decays $\Upsilon(10753)\to \eta_b \omega(\eta^{(\prime)})$. Similar to the early treatment~\cite{2021liPRD104-034036,2022liPRD105-114041,2022baiPRD105-074007}, we also interpret the $\Upsilon(10753)$ as a $4S$-$3D$ mixed state~\cite{2021liPRD104-034036,2022liPRD105-114041,2022baiPRD105-074007}. A short discussion of the $4S$-$3D$ mixing scheme for the $\Upsilon(10753)$ from physical point of view is given in Refs. \cite{2010badalianPAN73-138,2021liPRD104-034036,2009badalianPRD79-037505}. In terms of the $S$-$D$ mixing scheme, the wave function of the $\Upsilon(10753)$ is given by~\cite{2010badalianPAN73-138,2021liPRD104-034036}
\begin{equation}\label{eq:wf10753}
	\tilde{\Upsilon}(10753) = \tilde{\Upsilon}(4S) \sin\theta  + \tilde{\Upsilon}_1(3\,{}^3D_1) \cos\theta\,,
\end{equation}
where $\theta$ is a mixing angle to describe the proportion of the partial waves. $\tilde{\Upsilon}(4S)$ and $\tilde{\Upsilon}_1(3\,{}^3D_1)$ describe the wave functions of the pure $\Upsilon(4S)$ and $\yd$ states, respectively. After taking into account the $S$-$D$ mixing, the $\Upsilon(10580)$ that was usually regraded as the $4S$ state \cite{2022particledatagroupPoTaEP2022-083C01} should be currently interpreted as another $4S$-$3D$ mixture, and accordingly has the wave function
\begin{equation}\label{eq:wfY10580}
	\tilde{\Upsilon}(10580) = \tilde{\Upsilon}(4S) \cos\theta  - \tilde{\Upsilon}(3\,{}^3D_1) \sin\theta\,.
\end{equation}
The mixing angle can be obtained from the dielectron decay width~\cite{2010badalianPAN73-138,2021liPRD104-034036}
\begin{eqnarray}\label{eq:gammaee}
	\Gamma_{ee} &=& \frac{4\pi e_b^2\alpha^2}{3m_{\Upsilon(10580)}}f_\Upsilon^2\beta_V \nonumber \\
	&=& \frac{4e_b^2\alpha^2}{m_{\Upsilon(10580)}^2}\big|\tilde{\Upsilon}(10580)(0)\big|^2\xi\beta_V .
\end{eqnarray} 
Here $f_\Upsilon$ stands for the decay constant, $e_b=-1/3$ the charge of the $b$ quark in units of $|e|$, $\alpha = 1/137$ the fine structure constant, $m_{\Upsilon(10580)}=10579.4~\mathrm{MeV}$ the mass of the state $\Upsilon(10580)$, $\tilde{\Upsilon}(10580)(0)$ the wave function of the $\Upsilon(10580)$ at the origin. In addition, $\xi=0.968$ is a relativistic factor and $\beta_V=0.80\pm 0.01$ describes the QCD one-loop perturbative corrections~\cite{2010badalianPAN73-138}. In view of $\tilde{\Upsilon}(4S)(0)=1.506~\mathrm{GeV^{3/2}}$ and $\tilde{\Upsilon}_1(3\,{}^3D_1)(0)=0.0956 ~\mathrm{GeV^{3/2}}$~\cite{2010badalianPAN73-138} together with Eq.~\eqref{eq:wfY10580}, and with the well established $\Gamma_{ee}(\Upsilon(10580))=(0.272\pm 0.029)~\mathrm{keV}$ \cite{2022particledatagroupPoTaEP2022-083C01} for the physical states $\Upsilon(10580)$, one finds the mixing angle $\theta = (27.7\pm 5.0)^\circ$, where the error is from the uncertainty of the dielectron decay width for the $\Upsilon(10580)$. This estimation is consistent with the recent predicted value of $(33\pm 4)^\circ$ by Li et al~\cite{2021liPRD104-034036}.

Another way to estimate the mixing angle $\theta$ is to use the mass-mixing formula \cite{2022particledatagroupPoTaEP2022-083C01}. In terms of the quadratic mass mixing scheme, the mixing angle $\theta$ is governed by
\begin{equation}
	\cos^2\theta = \frac{m_{\Upsilon(10753)}^2-m_{\Upsilon(4S)}^2}{m_{\Upsilon(10753)}^2-m_{\Upsilon(10580)}^2}\,.
\end{equation}
A search in the literature yields that the mass of the pure $\Upsilon(4S)$ were theoretically predicted to range from $10607~\mathrm{MeV}$ to $10635~\mathrm{MeV}$~\cite{2018wangEPJC78-915,2015godfreyPRD92-054034,2016segoviaPRD93-074027,1985godfreyPRD32-189,2010badalianPAN73-138,2009badalianPRD79-037505}. This mass range leads to the mixing angle between $23.4^\circ$ and $36.1^\circ$, agreeing with the results obtained by the foregoing fitting procedure. Moreover, such mass range of the $\Upsilon(4S)$ would require the mass of the pure $\yd$ to be $10698$--$10725~\mathrm{MeV}$, coinciding also with the early theoretically predicted masses from $10653$ to $10717~\mathrm{MeV}$~\cite{2018wangEPJC78-915,2015godfreyPRD92-054034,2016segoviaPRD93-074027,1985godfreyPRD32-189,2010badalianPAN73-138,2009badalianPRD79-037505}. These estimations provide ostensible feasibility of the formation of the $\Upsilon(10753)$ by the $4S$-$3D$ mixing, which needs verification by comparing various predictions of this mixing scheme to the experimental results. Furthermore, it also indicates that the $\yd$ is the dominant component to form the $\Upsilon(10753)$.

In the following, we shall, based on the $4S$-$3D$ mixing scenario, investigate the decays of the $\Upsilon(10753)$ to the $\eta_b (1S)$ with emission of the $\omega$, $\eta$, and $\eta^\prime$.

\subsection{The Effective Lagrangians}

According to the Heavy Quark Effective Theory (HQET), the interactions of the $S$-wave bottomonium $\Upsilon(nS)$ and $\eta_b(nS)$ with the bottom and anti-bottom mesons are described by the Lagrangian~\cite{1997casalbuoniPR281-145,2021wuPRD104-074011,2021liPRD104-034036,2022liPRD105-114041}
\begin{align}\label{eq:LS}
	\mathcal{L}_S &= \ii g_{\Upsilon BB}\Upsilon_\mu \bar{B}^\dagger \pararrowk{\mu} B^\dagger \nonumber\\
	& + g_{\Upsilon BB^*} \epsilon_{\mu\nu\alpha\beta} \partial^\mu \Upsilon^\nu (\bar{B}^{*\alpha\dagger}\pararrowk{\beta} B^\dagger- \bar{B}^\dagger\pararrowk{\beta} B^{*\alpha\dagger})\nonumber\\
	&+\ii  g_{\Upsilon B^*B^*} \Upsilon_\mu (\bar{B}^{*\dagger}_\nu \pararrowk{\nu}B^{*\mu\dagger} + \bar{B}^{*\dagger}_\mu \pararrowk{\nu} B^{*\dagger}_\nu - \bar{B}^{\dagger}_\nu \pararrow_\mu B^{*\nu\dagger})\nonumber\\
	&-\ii g_{\eta_bBB^*} \eta_b(\bar{B}^\dagger\pararrowk{\mu}B^{*\dagger}_\mu + \bar{B}^{*\dagger}_\mu\pararrowk{\mu}B^\dagger)\nonumber\\
	&- g_{\eta_bB^*B^*} \epsilon_{\mu\nu\alpha\beta} \partial^\mu \eta_b \bar{B}^{*\nu\dagger}\pararrowk{\alpha}B^{*\beta\dagger} + \mathrm{H.c.}\,.
\end{align}
Here $B^{(*)} = (B^{(*)0}\,,B^{(*)+}\,,B_s^{(*)0})$ and $\bar{B}^{(*)} = (\bar{B}^{(*)0}\,,B^{(*)-}\,,\bar{B}_s^{(*)0})$. The coupling constants are linked to each other by a global constant $g_1$ and the mass of the involved mesons, i.e.,
\begin{subequations}\label{eq:YBB}
	\begin{align}
		g_{\Upsilon BB} &= 2g_1m_B \sqrt{m_\Upsilon}\,,\\
		g_{\Upsilon B^*B^*} &= 2g_1m_{B^*} \sqrt{m_\Upsilon}\,,\\
		g_{\Upsilon BB^*} &= 2g_1 \sqrt{ m_{B^*} m_{B}/m_\Upsilon}\,,\\
		g_{\eta_bBB^*} &= 2g_1 \sqrt{ m_{B^*} m_{B} m_{\eta_b}}\,,\\
		g_{\eta_bB^*B^*} &=2g_1m_{B^*}/\sqrt{m_{\eta_b}}\,.
	\end{align}
\end{subequations}
The value of $g_1$ could be determined by the experimental or theoretical branching ratios of the open $b$-flavored strong decays. For those below the $B\bar{B}$ threshold, the $g_1$ could be calculated by~\cite{2015liPRD91-034020,2018huangPRD98-054008}
\begin{equation}
	g_1 = \frac{m_\Upsilon}{2m_B f_\Upsilon},
\end{equation}
with the decay constant $f_\Upsilon$, which could be extracted from the dielectron width $\Gamma_{ee}$ using Eq. \eqref{eq:gammaee}.

The coupling of the $D$-wave bottomonium $\yd$ to a pair of bottom and anti-bottom mesons is ~\cite{1997casalbuoniPR281-145,2021liPRD104-034036,2022liPRD105-114041}
\begin{align}\label{eq:LD}
	\mathcal{L}_D &= \ii g_{\Upsilon_1 BB}{\Upsilon_1}_\mu \bar{B}^\dagger \pararrowk{\mu} B^\dagger -\ii  g_{\Upsilon_1 B^*B^*} {\Upsilon_1}_\mu (\bar{B}^{*\dagger}_\nu \partial^\nu B^{*\mu\dagger} \nonumber \\
	&-4\bar{B}^{*\dagger}_\nu \pararrowk{\mu}B^{*\nu\dagger} - B^{*\nu\dagger} \partial_\nu  \bar{B}^{*\mu\dagger})\nonumber\\
	& - g_{\Upsilon_1 BB^*} \epsilon_{\mu\nu\alpha\beta} \partial^\nu \Upsilon_1^\alpha (\bar{B}^{*\beta\dagger}\pararrowk{\mu} B^\dagger- \bar{B}^\dagger\pararrowk{\mu} B^{*\beta\dagger})\,,
\end{align}
where the coupling constants are governed by another global factor $g_2$ in the following form
\begin{subequations}\label{eq:Y1BB}
	\begin{align}
		g_{\Upsilon_1 BB}& = \frac{10}{\sqrt{15}}g_2 m_B\sqrt{m_{\Upsilon_1}}\,,\label{eq:Y1BBa}\\
		g_{\Upsilon_1 BB^*}&=\frac{5}{\sqrt{15}}g_2\sqrt{m_Bm_{B^*}/m_{\Upsilon_1}}\,,\label{eq:Y1BBb}\\
		g_{\Upsilon_1 B^*B^*}&=\frac{1}{\sqrt{15}}g_2m_{B^*}\sqrt{m_{\Upsilon_1}}\,.\label{eq:Y1BBc}
	\end{align}
\end{subequations}

Based on the heavy quark limit and chiral symmetry, the interactions of the light vector and pseudoscalar mesons with the heavy bottom mesons read \cite{2021wuEPJC81-193,2022wangPRD106-074015,2023wuPRD107-034028}
\begin{eqnarray}\label{eq:LVP}
	\mathcal{L} &=& -\ii g_{BBV} B_i^\dagger \pararrowk{\mu}B^{j}(V_\mu^\dagger)_j^i \nonumber\\
	&-&2 f_{B^*BV}\epsilon_{\mu\nu\alpha\beta} (\partial^\mu V^{\nu\dagger})_j^i(B_i^\dagger \pararrowk{\alpha}B^{*\beta j} \nonumber\\
	&-& B_i^{*\beta\dagger}\pararrowk{\alpha}B^{j})
	+\ii g_{B^*B^*V}B_i^{*\nu\dagger}\pararrowk{\mu}B_\nu^{*j}(V_\mu^\dagger)_j^i\nonumber \\
	&+& \ii 4 f_{B^*B^*V} B_{i\mu}^{*\dagger}(\partial^\mu V^{\nu\dagger} - \partial^\nu V^{\mu\dagger})_j^i B_\nu^{*j}\nonumber\\
	&-&\ii g_{B^*BP}\big(B^{i \dagger}\partial^{\mu} P_{ij}^\dagger B_\mu^{*j} - B_\mu^{*i\dagger}\partial^\mu P_{ij}^\dagger B^j\big) \nonumber\\
	&+& \frac{1}{2} g_{B^*B^*P}\epsilon_{\mu\nu\alpha\beta} B_i^{*\mu\dagger}\partial^\nu P^{ij\dagger}\pararrowk{\alpha} B_j^{*\beta} + \mathrm{H.c.}\,,
\end{eqnarray}
where the $V$ and $P$ are, respectively, the nonet vector and pseudoscalar mesons in the matrix form
\begin{subequations}\label{eq:vmatrix}
	\begin{align}
		V &= 
		\begin{pmatrix}
			\frac{\rho^0}{\sqrt{2}}+\frac{\omega}{\sqrt{2}}&\rho^+&K^{*+}\\
			\rho^-&-\frac{\rho^0}{\sqrt{2}}+\frac{\omega}{\sqrt{2}}&K^{*0}\\
			K^{*-}&\bar{K}^{*0}&\phi
		\end{pmatrix}\,,\\
		P &= \begin{pmatrix}
			\frac{\pi^0}{\sqrt{2}} + \frac{\beta \eta + \gamma \eta'}{\sqrt{2}} & \pi^+ & K^+\\
			\pi^- & -\frac{\pi^0}{\sqrt{2}} + \frac{\beta \eta + \gamma \eta'}{\sqrt{2}} & K^0\\
			K^- & \bar{K}_0 & - \gamma \eta + \beta \eta' \label{eq:P}
		\end{pmatrix}\,.
	\end{align}
\end{subequations}
Here $\beta = \cos(\theta_\mathrm{P} + \arctan\sqrt{2})$ and $\gamma = \sin(\theta_\mathrm{P} + \arctan\sqrt{2})$ with the $\eta$-$\eta'$ mixing angle $\theta_\mathrm{P}$ ranging from $-24.6^\circ$ to $-11.5^\circ$ \cite{2022particledatagroupPoTaEP2022-083C01,2013liPRD88-014010,2023wuPRD107-034028,2012wangPRD85-074015,1996amslerPRD53-295,1983rosnerPRD27-1101}. 

The coupling constants $g_{B^{(*)}B^{(*)}V}$ and $g_{B^{(*)}B^{(*)}P}$ could be determined using the following relations \cite{2021wuEPJC81-193}
\begin{subequations}\label{eq:gddvs}
	\begin{align}
		g_{BBV} &= g_{B^*B^*V} = \frac{\beta g_V}{\sqrt{2}}\,,\label{eq:gddvgdsdsv}\\
		f_{B^*BV}&= \frac{f_{B^*B^*V}}{m_{B^*}} = \frac{\lambda g_V}{\sqrt{2}}\,,\label{eq:fdsdvfdsdsv}\\
		g_{B^*B^*P} &= \frac{g_{B^*BP}}{\sqrt{m_Bm_{B^*}}}= \frac{2g}{f_{\pi}}\,.\label{eq:gddp}
	\end{align}
\end{subequations}
Here $\beta=0.9$ and $g_V = m_\rho / f_\pi$ with the pion decay constant $f_\pi = 132~\mathrm{MeV}$ \cite{1997casalbuoniPR281-145} and the $\rho$ meson mass $m_{\rho}=775.26~\mathrm{MeV}$~\cite{2022particledatagroupPoTaEP2022-083C01}. Moreover, $\lambda=0.56~\mathrm{GeV^{-1}}$ and $g = 0.59$ based on the matching of the form factors obtained from the light cone sum rule and from the lattice QCD calculations \cite{2003isolaPRD68-114001}.

\subsection{Transition Amplitudes}
According to the mixed wave function of the $\Upsilon(10753)$ in Eq. \eqref{eq:wf10753}, the amplitude of the decays we consider is written as
\begin{equation}
	\mathcal{M} = \mathcal{M}^S\sin\theta + \mathcal{M}^D\cos\theta\,,
\end{equation}
where $\mathcal{M}^S$ and $\mathcal{M}^D$ are the amplitudes due to the pure $\Upsilon(4S)$ and $\yd$ contributions, respectively, of which the proportion is described by the mixing angle $\theta$. The amplitude $\mathcal{M}_\omega$ for the $\Upsilon(10753)\to \eta_b \omega$ and $\mathcal{M}_P$ for the $\Upsilon(10753)\to \eta_b P$ ($P=\eta\,,\eta^\prime$) are, respectively, given as
\begin{subequations}\label{eq:msd}
	\begin{align}
		\mathcal{M}^{S(D)}_\omega &= \frac{1}{\sqrt{2}} \varepsilon^\mu(\Upsilon) \varepsilon^{*\nu}(\omega)(N^{S(D)}_{\mu\nu} + C^{S(D)}_{\mu\nu})\,,\\
		\mathcal{M}^{S(D)}_P &= \varepsilon^\mu(\Upsilon) \big[x(N^{S(D)}_\mu+C^{S(D)}_\mu) +yS^{S(D)}_\mu\big]\,,
	\end{align}
\end{subequations}
where $\varepsilon^\mu(\Upsilon)$ and $\varepsilon^{*\nu}(\omega)$ are the polarization vectors of the $\Upsilon(4S)/ \yd$ and $\omega$, respectively. The factors $x$ and $y$ equal to $\beta/\sqrt{2}$ and $-\gamma$ for the $\Upsilon(10753)\to \eta_b\eta$ and to $\gamma/ \sqrt{2}$ and $ \beta$ for the $\Upsilon(10753)\to \eta_b\eta^\prime$. The tensor and vector structures $C_{\mu(\nu)}$'s, $N_{\mu(\nu)}$'s, and $S_{\mu}$ correspond to the summation of the charged, neutral, and strange bottom meson loop integrals, governed by the foregoing Lagrangians in Eqs. \eqref{eq:LS}, \eqref{eq:LD}, and \eqref{eq:LVP}. Due to the light flavor symmetry, the charged, neutral, and strange loop integrals have the same form, for which only masses of the intermediate mesons are different. 

Selecting the neutral loop in Fig. \ref{fig:feyndiags}(a) as an example, the tensor structures $N_{\mu\nu}^{S(D)}$ for the case of $\Upsilon(p)\to B^{*0}(p_1)\bar{B}^{*0}(p_2)[B^0(q)]\to \eta_b(p_4)\omega(p_3)$ explicitly read
\begin{widetext}
	\begin{subequations}
		\begin{align}
			N_{\mu\nu}^{S(a)} = \ii^3\int \frac{\dd[4]{q}}{(2\pi)^4}& \big[-g_{\Upsilon B^* B^*}\big(p_{1\sigma}g_{\mu\alpha}-(p_1-p_2)_\mu g_{\alpha\sigma}-p_{2\alpha}g_{\mu\sigma}\big)\big]\big[-2f_{B^*BV}p_3^\lambda(p_1+q)^\delta \epsilon_{\lambda \nu \delta \beta}\big]\nonumber\\
			\times&[-g_{\eta_bBB^*}(q-p_2)_\rho] D^{\alpha\beta}(p_1\,,m_{B^{*0}})D^{\sigma\rho}(p_2\,,m_{\bar{B}^{*0}})D(q\,,m_{B^0})F(q\,,m_{B^0})\,,\\
			N_{\mu\nu}^{D(a)} = \ii^3\int \frac{\dd[4]{q}}{(2\pi)^4}& \big[-g_{\Upsilon_1 B^* B^*}\big(p_{1\sigma}g_{\mu\alpha}-4(p_1-p_2)_\mu g_{\alpha\sigma}-p_{2\alpha}g_{\mu\sigma}\big)\big]\big[-2f_{B^*BV}p_3^\lambda(p_1+q)^\delta \epsilon_{\lambda \nu \delta \beta}\big]\nonumber\\
			\times &[-g_{\eta_bBB^*}(q-p_2)_\rho] D^{\alpha\beta}(p_1\,,m_{B^{*0}})D^{\sigma\rho}(p_2\,,m_{\bar{B}^{*0}})D(q\,,m_{B^0})F(q\,,m_{B^0})\,.
		\end{align}
	\end{subequations}
	Additionally, the vector structures $N_{\mu}^{S(D)}$ for the case of $\Upsilon(p)\to B^{*0}(p_1)\bar{B}^{*0}(p_2)[B^0(q)]\to \eta_b(p_4)\eta^{(\prime)}(p_3)$ are
	\begin{subequations}
		\begin{align}
			N_{\mu}^{S(a)} =\ii^3\int \frac{\dd[4]{q}}{(2\pi)^4}& \big[-g_{\Upsilon B^* B^*}\big(p_{1\sigma}g_{\mu\alpha}-(p_1-p_2)_\mu g_{\alpha\sigma}-p_{2\alpha}g_{\mu\sigma}\big)\big]\big[-g_{B^*BP}p_{3\beta}\big]\big[-g_{\eta_b BB^*} (q-p_2)_\rho\big]\nonumber\\
			\times& D^{\alpha\beta}(p_1\,,m_{B^{*0}})D^{\sigma\rho}(p_2\,,m_{\bar{B}^{*0}})D(q\,,m_{B^0})F(q\,,m_{B^0})\,,\\
			N_{\mu}^{D(a)} =\ii^3\int \frac{\dd[4]{q}}{(2\pi)^4}& \big[-g_{\Upsilon_1 B^* B^*}\big(p_{1\sigma}g_{\mu\alpha}-4(p_1-p_2)_\mu g_{\alpha\sigma}-p_{2\alpha}g_{\mu\sigma}\big)\big]\big[-g_{B^*BP}p_{3\beta}\big]\big[-g_{\eta_b BB^*} (q-p_2)_\rho\big]\nonumber\\
			\times& D^{\alpha\beta}(p_1\,,m_{B^{*0}})D^{\sigma\rho}(p_2\,,m_{\bar{B}^{*0}})D(q\,,m_{B^0})F(q\,,m_{B^0})\,.
		\end{align}
	\end{subequations}
\end{widetext}
Here $D$ and $D^{\mu\nu}$, respectively, represent the propagators for the scalar $B$ and vector $B^*$ in the following form
\begin{subequations}
	\begin{align}
		D(p,m_B) &= \frac{1}{p^2-m_B^2+\ii \epsilon}\,,\\
		D^{\mu\nu}(p,m_{B^*}) &= \frac{-g^{\mu\nu} + p^\mu p^\nu/m_{B^*}^2}{p^2-m^2_{B^*}+\ii \epsilon}\,.
	\end{align}
\end{subequations}
Moreover, the $F(q\,,m_{B^0}) $ is a form factor to account for the off-shell effect as well as the inner structure of the exchanged mesons~\cite{1994tornqvistZPC-PaF61-525,1994locherZPA-HaN347-281,1996gortchakovZPA-HaN353-447,1997liPRD55-1421,2005chengPRD71-014030}. Here, we adopt a dipole form factor~\cite{2022wangPRD106-074015}
\begin{equation}\label{eq:formfactor}
	F(q,m) = \left(\frac{m^2-\Lambda^2}{q^2-\Lambda^2}\right)^2\,,
\end{equation}
where $q$ and $m$ stand for the momentum and mass of the exchanged meson, respectively; $\Lambda = m + \alpha \Lambda_{\mathrm{QCD}}$ with $\Lambda_{\mathrm{QCD}} = 0.22~\mathrm{GeV}$~\cite{2005chengPRD71-014030}, in which the $\alpha$ is usually regarded as an undetermined parameter in the vicinity of unity. In present calculations, we take the $\alpha$ ranging from $0.2$ to $1.0$.
To obtain the charged and strange terms $C_{\mu(\nu)}$ and $S_{\mu}$ in Eq.\eqref{eq:msd}, we only need to replace the neutral $B$'s by charged and strange ones. The tensor and vector structures $N_{\mu(\nu)}$ for the Figs. \eqref{fig:feyndiags}(b)--(f) are given in the Appendix \ref{app:amps}. The partial decay widths of the $\Upsilon(10753) \to \eta_b \omega (\eta^{(\prime)})$ are given by
\begin{equation}\label{eq:partialwidth}
	\Gamma[\Upsilon(10753) \to \eta_b \omega (\eta^{(\prime)})] = \frac {|\vec{p}||\mathcal{M}_{\Upsilon(10753) \to \eta_b \omega (\eta^{(\prime)})}|^2} {24\pi m^2_{\Upsilon(10753)}}\,,
\end{equation}
where a summation over the polarizations of initial and final vector mesons is included. $\vec{p}$ is the three momentum of the final light mesons.

\section{Numerical Results and Discussion}
\label{sec:results}

In Ref. \cite{2018wangEPJC78-915}, the total width of the pure $\Upsilon(4S)$ was predicted to be $24.7~\mathrm{MeV}$, of which the branching fraction for the $\Upsilon(4S)\to B\bar{B}$ approaches unity. In view of the experimental measurement that the branching fractions for the $\Upsilon(10580)\to B^0\bar{B}^0$ and $\Upsilon(10580)\to B^+B^-$ are nearly equal and their sum is larger than 0.96~\cite{2022particledatagroupPoTaEP2022-083C01}, we therefore assume that the total width of the pure $\Upsilon(4S)$ are saturated by the two processes $\Upsilon(4S)\to B^0\bar{B}^0$ and $\Upsilon(4S)\to B^+B^-$ with equal proportion. According to the two-body decay model together with the interactions in Eq. \eqref{eq:LS}, we get the coupling constant of the $\Upsilon(4S)$ to $B\bar{B}$ is about $13.343$, and $g_1=0.388~\mathrm{GeV^{-3/2}}$. Then, the other relevant constants can be obtained using Eq. \eqref{eq:YBB}.

To determine the coupling constants for the interactions of the pure $\yd$ with the $B^{(*)}_{(s)}\bar{B}^{(*)}_{(s)}$, we employ the theoretically predicted partial widths of the $\yd$ decaying to $B^{*}\bar{B}^{(*)}$ \cite{2018wangEPJC78-915}. We obtain the coupling constants of the $\yd$ to the non-strange $B\bar{B}$, $B\bar{B}^*+\mathrm{c.c.}$, and $B^*\bar{B}^{*}$ are $3.492$, $0.394~\mathrm{GeV^{-1}}$, and $4.215$, respectively. It is worth mentioning that these above values go against the relations in Eq.\eqref{eq:Y1BB}. Consequently, in our calculations we ignore the constraint among Eqs. \eqref{eq:Y1BBa}, \eqref{eq:Y1BBb}, and \eqref{eq:Y1BBc} and use three different $g_2$'s instead, namely, for the coupling of the $\yd$ to $B\bar{B}$, $B\bar{B}^*+\mathrm{c.c.}$, and $B^*\bar{B}^{*}$ and the corresponding strange bottom meson pairs, the global constants $g_2$'s equal to $0.078~\mathrm{GeV^{-3/2}}$, $0.188~\mathrm{GeV^{-3/2}}$, and $0.938~\mathrm{GeV^{-3/2}}$, respectively.

In order to estimate the coupling constant of the $\eta_b$ and the possible bottom meson pairs, we need to calculate the decay constant of the $\Upsilon(1S)$ using Eq. \eqref{eq:gammaee}. In terms of $\Gamma_{ee} = 1.340~\mathrm{keV}$ for the $\Upsilon(1S)$ \cite{2022particledatagroupPoTaEP2022-083C01}, $f_{\Upsilon}$ is about $0.715~\mathrm{GeV}$, and thereby leading to $g_1 = 0.407~\mathrm{GeV^{-3/2}}$, which is similar to the result in Ref. \cite{2018huangPRD98-054008}. For easy reference, we summarized the global factors $g_1$ and $g_2$ obtained above in Table \ref{tab:globalgg}.

\begin{table}[htbp]
	\caption{Global parameters $g_1$ and $g_2$ (Units: $\mathrm{GeV^{-3/2}}$) we employed in the calculations. Their estimations are based on the theoretical and experimental data in Refs. \cite{2018wangEPJC78-915,2022particledatagroupPoTaEP2022-083C01}.}
	\label{tab:globalgg}
	\begin{ruledtabular}
		\begin{tabular}{lccr}
		$g_1\,,g_2$	&$B_{(s)}\bar{B}_{(s)}$& $B_{(s)}\bar{B}_{(s)}^*+\mathrm{c.c.}$&$B_{(s)}^*\bar{B}^*_{(s)}$\\
			\colrule
			$\Upsilon(4S)$ & 0.388 & 0.388 & 0.388\\
			$\yd$ & 0.078 & 0.188 & 0.938\\
			$\eta_b$ & $\cdots$ & 0.407 & 0.407\\
		\end{tabular}
	\end{ruledtabular}
\end{table}

\begin{figure}[htbp]
	\centering
	\includegraphics[width=0.94\linewidth]{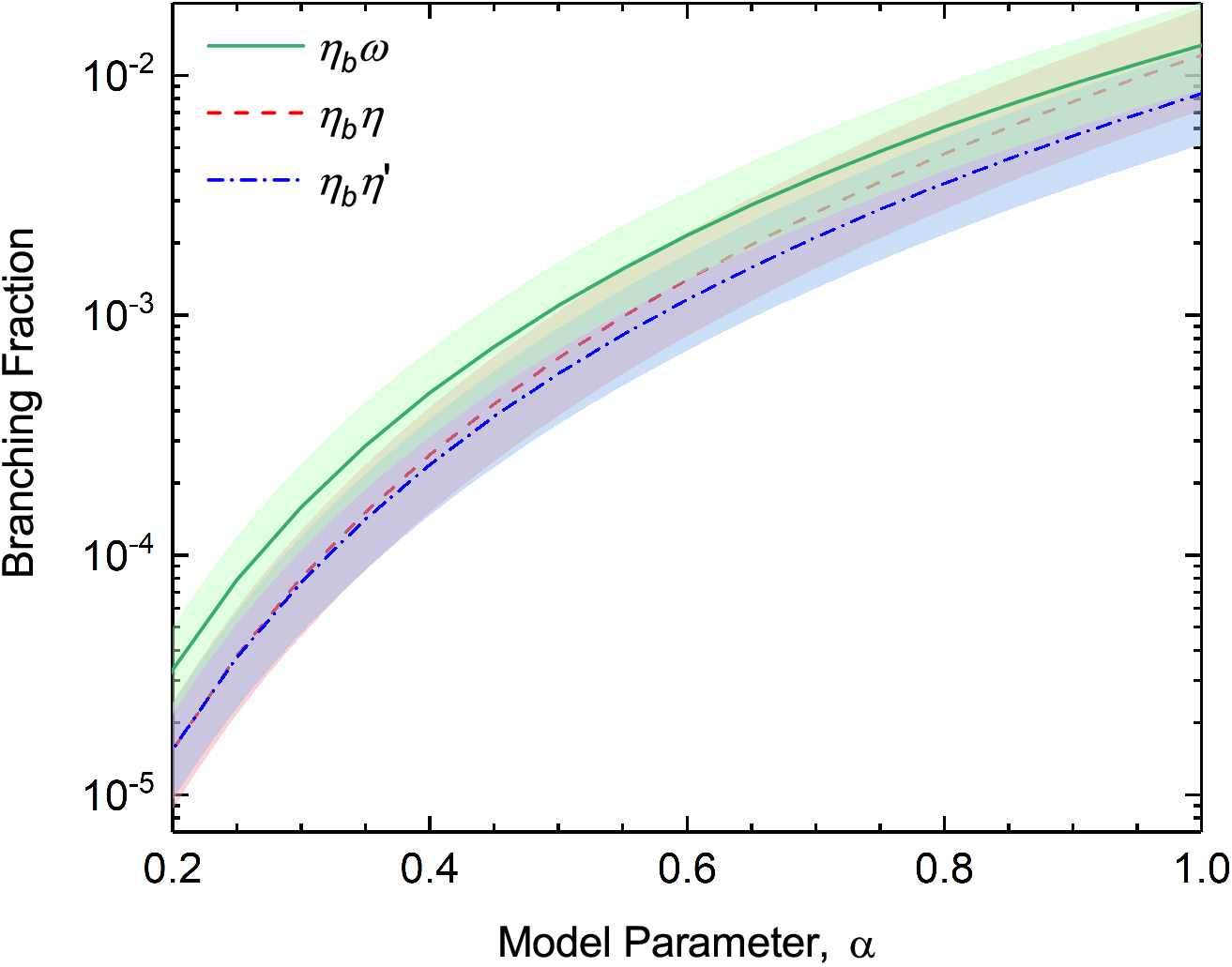}
	\caption{(Color online) Branching fractions of the processes $\Upsilon(10753) \to \eta_b \omega$, $\eta_b\eta$, and $\eta_b\eta^\prime$. The $4S$-$3D$ mixing angle is fixed to be $33^\circ$ and the $\eta$-$\eta^\prime$ mixing angle is taken to be the widely used value of $-19.1^\circ$ that was determined by DM2 Collaboration~\cite{1990dm2collaborationPRD41-1389}. The light-colored bands indicate the margin of error resulting mainly from the errors of the mass and width for the $\Upsilon(10753)$ and of the $4S$-$3D$ mixing angle.}
	\label{fig:bromgetas}
\end{figure}

Figure \ref{fig:bromgetas} shows the branching fractions for the decays of $\Upsilon(10753)\to \eta_b\omega$, $\eta_b\eta$, and $\eta_b\eta'$ as a function of the model parameter $\alpha$ introduced in the form factor (see Eq. \eqref{eq:formfactor}). The calculation were performed using the $4S$-$3D$ mixing angle of $33^\circ$ and the $\eta$-$\eta^\prime$ mixing angle of $-19.1^\circ$ that was determined by DM2 Collaboration~\cite{1990dm2collaborationPRD41-1389}. The parameter $\alpha$ is varied from $0.2$ to $1.0$. It is seen that the results are strongly sensitive to the parameter $\alpha$, changing from about $10^{-5}$ to $10^{-2}$. Explicitly,
\begin{equation*}
	\begin{split}
		\mathcal{B} (\Upsilon(10753)\to \eta_b\omega) = 3.3\times 10^{-5} \sim 1.3\times 10^{-2}\,,\\
		\mathcal{B} (\Upsilon(10753)\to \eta_b\eta) = 1.5\times 10^{-5} \sim 1.2\times 10^{-2}\,,\\
		\mathcal{B} (\Upsilon(10753)\to \eta_b\eta') = 1.5\times 10^{-5} \sim 8.4\times 10^{-3}\,.
	\end{split}
\end{equation*}

In the absence of relevant experimental data, it seems difficult to narrow down the $\alpha$ range. However, it is recalled that the $\Upsilon(10753)$ have the same quantum numbers $J^{PC} = 1^{--}$ with $\Upsilon(10580)$ and $\Upsilon(10860)$, and its mass is between $m_{\Upsilon(10580)}$ and $m_{\Upsilon(10860)}$. Within the $S$-$D$ mixing framework, it is, hence, plausible to expect that the $\Upsilon(10753)$ has comparable decay modes to the $\Upsilon(10580)$ and $\Upsilon(10860)$. In view of the fact that the branching fraction for the $\Upsilon(10580)\to \eta_b\omega$ was measured to be less than $1.8\times 10^{-4}$ and for the $\Upsilon(10860)\to \eta_b\omega$ it was smaller than $1.3\times 10^{-3}$~\cite{2022particledatagroupPoTaEP2022-083C01}, we could limit the $\alpha$ value to $0.3\sim 0.5$. Such small $\alpha$'s or even smaller ones were selected in previous work~\cite{2022liPRD105-114041,2023wuPRD107-034028}. In this range of the $\alpha$, the predicted partial widths for all the decays we consider are between 1 and 50 keV:
\begin{equation*}
	\begin{split}
		\Gamma (\Upsilon(10753)\to \eta_b\omega) = (5.6\sim 38.8)~\mathrm{keV}\,,\\
		\Gamma (\Upsilon(10753)\to \eta_b\eta) = (2.8\sim 23.4)~\mathrm{keV}\,,\\
		\Gamma (\Upsilon(10753)\to \eta_b\eta') = (2.7\sim 20.2)~\mathrm{keV}\,.
	\end{split}
\end{equation*}

It should be noted that the partial decay widths of the $\Upsilon(10753)\to \eta_b\omega$ we obtained using the $S$-$D$ mixing scenario are much smaller than those predicted when assigning the $\Upsilon(10753)$ as a tetraquark state, under which the predicted width is $(2.46^{+4.70}_{-1.60})~\mathrm{MeV}$~\cite{2019wangCPC43-123102}. This great difference is quite favorable for us to distinguish the internal structure of the $\Upsilon(10753)$ when we have relevant experimental data in the future. That is to say, when the future experiments give smaller widths on the order of keV for the $\Upsilon(10753)\to \eta_b\omega$, the $\Upsilon(10753)$ appears to favor the $4S$-$3D$ mixed state.

\begin{figure}[htbp]
	\centering
	\includegraphics[width=0.94\linewidth]{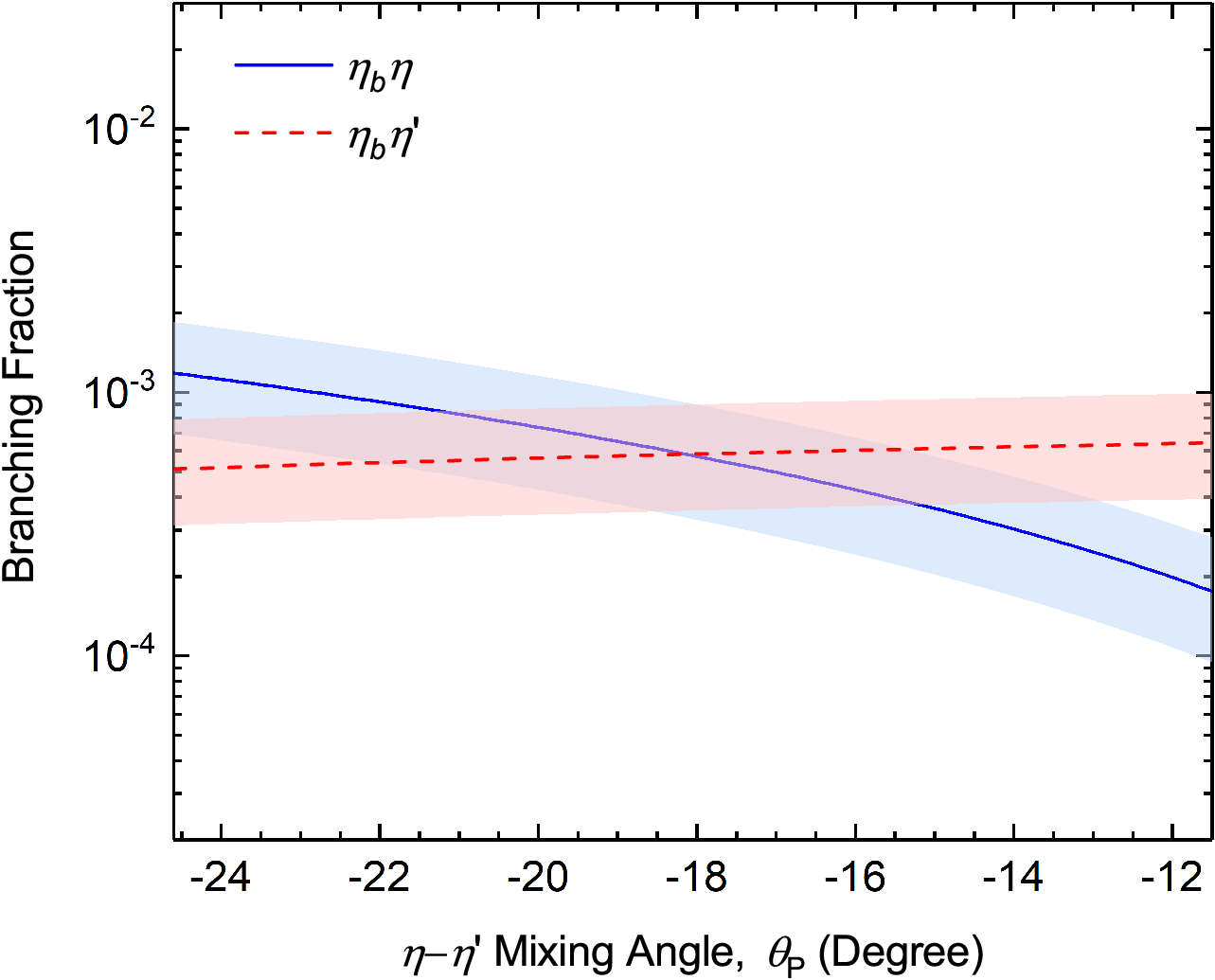}
	\caption{(Color online) Branching factions of the decays $\Upsilon(10753)\to \eta_b\eta$ and $\Upsilon(10753)\to\eta_b\eta^\prime$ as a function of the $\eta$-$\eta'$ mixing angle $\theta_{\mathrm{P}}$. The model parameter $\alpha$ is fixed at 0.5. The light-colored bands describe the errors due to errors of the $\Upsilon(10753)$ mass and width, and the $4S$-$3D$ mixing angle. Again, the $4S$-$3D$ mixing angle is set to be $33^\circ$.}
	\label{fig:bretaetapvsmix}
\end{figure}

In Fig.~\ref{fig:bretaetapvsmix} the branching fractions of the decays $\Upsilon(10753) \to \eta_b\eta$ and $\eta_b\eta'$ are plotted for different $\eta$-$\eta'$ mixing angles. These calculations were performed at the fixed model parameter $\alpha = 0.5$. As seen, with increasing the $\eta$-$\eta'$ mixing angle $\theta_{\mathrm{P}}$, the branching fraction for the $\Upsilon(10753)\to \eta_b\eta$ decreases distinctly, while for the $\Upsilon(10753)\to \eta_b\eta^\prime$ the branching fraction exhibits only a slight increase. In the vicinity of $\theta_{\mathrm{P}}=-18^\circ$, the branching fractions for these two decays are more likely to be equal.

\begin{figure}[htbp]
	\centering
	\includegraphics[width=0.94\linewidth]{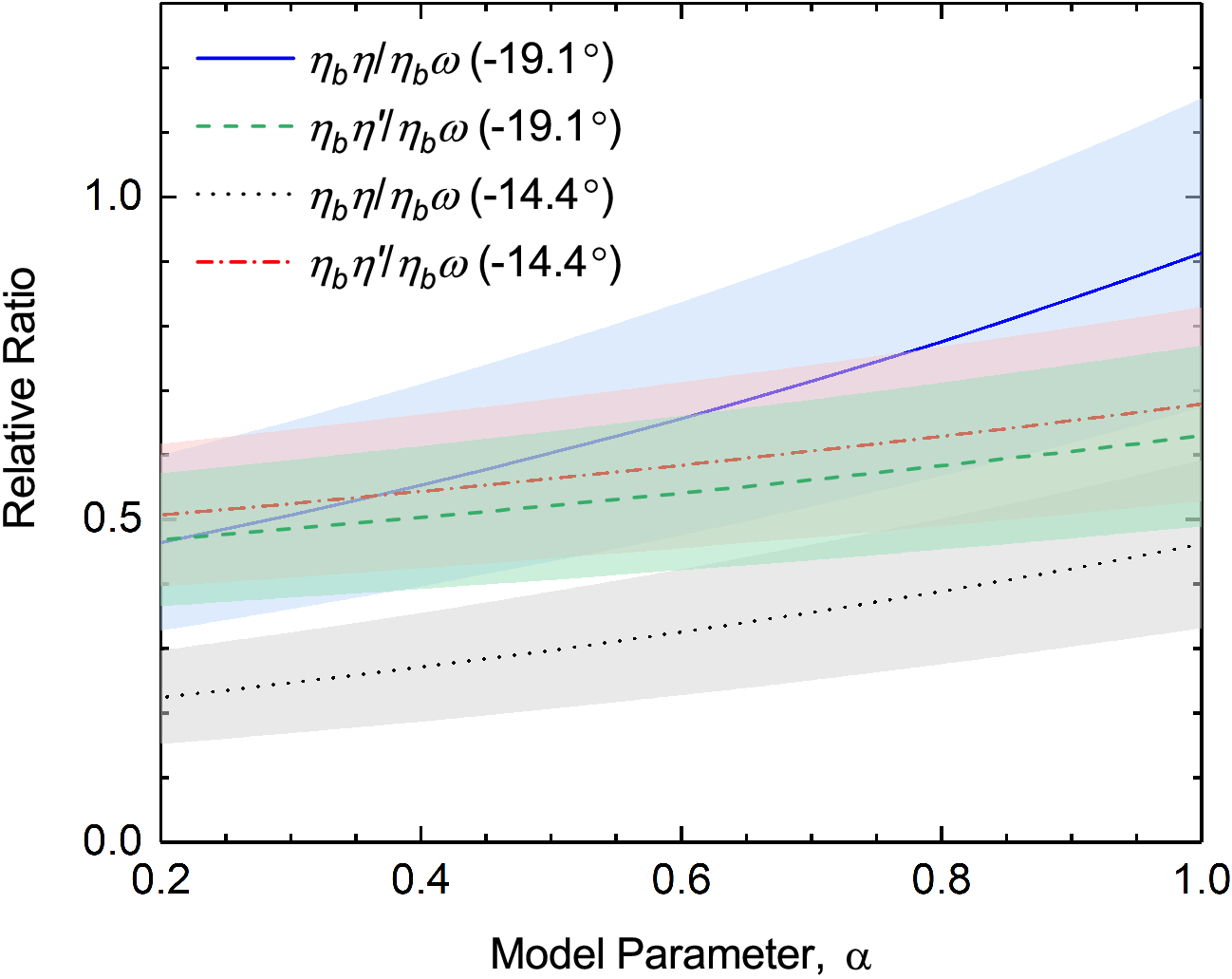}
	\caption{(Color online) Calculated ratios $R_{\eta/\omega}$ and $R_{\eta'/\omega}$ defined in Eq. \eqref{eq:Rs}. The light-colored bands depict the errors of the ratios that are caused by the errors of the $\Upsilon(10753)$ mass and the $4S$-$3D$ mixing angle. We select two $\eta$-$\eta'$ mixing angles of $-19.1^\circ$ \cite{1990dm2collaborationPRD41-1389} and $-14.4^\circ$ \cite{2009collaborationJHEP2009-105} and the $4S$-$3D$ mixing angle is $33^\circ$.}
	\label{fig:relratio}
\end{figure}

The strong $\alpha$ dependence is usually weakened when considering the relative ratios between the branching fractions of different processes. We here define the following ratios
\begin{subequations}\label{eq:Rs}
	\begin{align}
		R_{\eta/\omega} &= \frac{\mathcal{B} (\Upsilon(10753)\to \eta_b\eta)}{\mathcal{B} (\Upsilon(10753)\to \eta_b\omega)}\,,\\
		R_{\eta'/\omega} &= \frac{\mathcal{B} (\Upsilon(10753)\to \eta_b\eta')}{\mathcal{B} (\Upsilon(10753)\to \eta_b\omega)}\,.
	\end{align}
\end{subequations}
The calculated relative ratios for two $\eta$-$\eta^\prime$ mixing angles of $\theta_\mathrm{P} = -19.1^\circ$ and $\theta_\mathrm{P} = -14.4^\circ $ are shown in Fig.~\ref{fig:relratio}. It is seen that although the $\alpha$ dependence exists, it is actually weakened strongly in comparison to the absolute branching fractions shown in Fig. \ref{fig:bromgetas} that cover 3 orders of magnitude. On the other hand, the ratio $R_{\eta/\omega}$ changes strongly with the $\eta$-$\eta^\prime$ mixing angle, while the $R_{\eta^\prime/\omega}$ varies slightly. This finding is straightforward in view of the results shown in Fig. \ref{fig:bretaetapvsmix}.

\begin{figure}[htbp]
	\centering
	\includegraphics[width=0.94\linewidth]{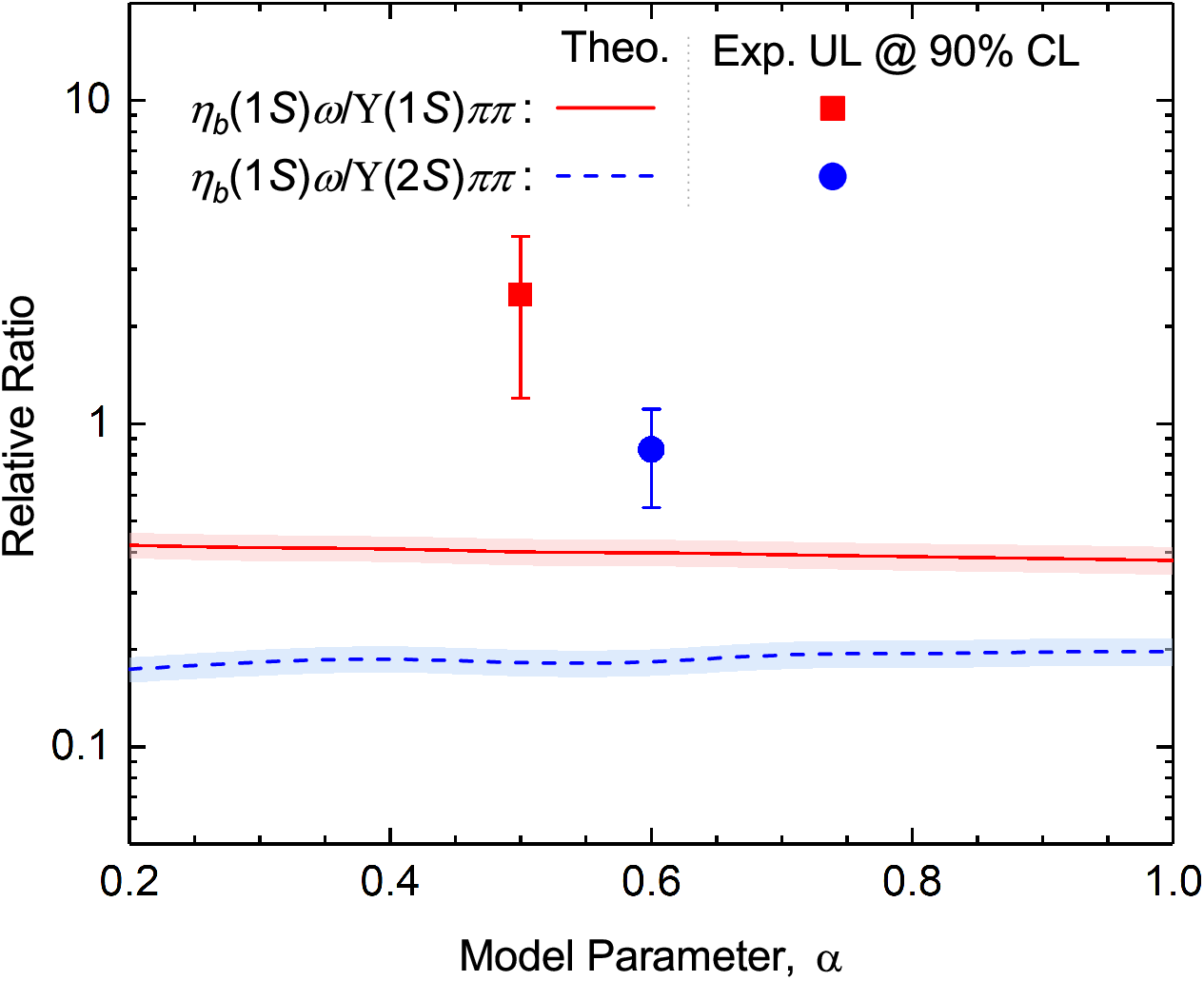}
	\caption{The ratio of $\mathcal{B} (\Upsilon(10753)\to \eta_b(1S)\omega)/\mathcal{B} (\Upsilon(10753)\to \Upsilon(nS)\pi^+\pi^-)$ ($n=1\,,2$) for different model parameter $\alpha$'s. The $4S$-$3D$ mixing angle is set to be $33^\circ$. As labeled by the texts in the graph, the lines represent the theoretical predictions and the points stand for the upper limits of the experimental results at 90\% confidence level, which are extracted from Refs. \cite{2019mizukJHEP2019-220,ceft}. The errors indicated by the light-colored band only consider the contribution from the $\Upsilon(10753)\to\eta_b\omega$.}
	\label{fig:omg2ypipi}
\end{figure}

Moreover, we calculated the ratio of $R_{\eta_b\omega/ (\Upsilon(nS)\pi\pi)} = \mathcal{B}(\Upsilon(10753)\to \eta_b(1S)\omega)/ \mathcal{B} (\Upsilon(10753)\to \Upsilon(nS)\pi^+\pi^-)$ ($n=1\,,2$). The calculation procedure for the $\Upsilon(10753)\to\Upsilon(nS)\pi^+\pi^-$ are similar with that by Bai et al in Ref. \cite{2022baiPRD105-074007} so that the related details are not repeated. The calculated results are shown in Fig. \ref{fig:omg2ypipi}. For comparison, we also show the upper limits of the experimental data at 90\% confidence level as the points, which are extracted from Refs. \cite{2019mizukJHEP2019-220,ceft}. It is clearly seen that the ratios $ R_{\eta_b\omega/\Upsilon(nS)\pi\pi} $ are nearly independent of the model parameter $\alpha$. The theoretical values of $R_{\eta_b\omega/\Upsilon(1S)\pi\pi} $ and $R_{\eta_b\omega/\Upsilon(2S)\pi\pi} $ are, respectively, around 0.4 and 0.2, which are in line with the experimental measurements with the upper limits of $2.5\pm 1.3$ and $0.83\pm 0.28$ for the cases of $\Upsilon(10753)\to\Upsilon(1S)\pi\pi$ and $\Upsilon(2S)\pi\pi$, respectively. This finding, to some extent, supports the interpretation of the $\Upsilon(10753)$ as a $4S$-$3D$ mixture.

\section{Summary} \label{sec:summary}

Calculations of partial decay widths of the $\Upsilon(10753)\to\eta_b\omega$, $\eta_b\eta$ and $\eta_b\eta^\prime$ through the intermediate meson loop mechanism have been performed using an effective Lagrangian approach. In the calculations, we assumed that the $\Upsilon(10753)$ is a $4S$-$3D$ mixed state with a mixing angle of $33^\circ$. The branching fractions of these decay processes are predicted to be $10^{-4}$--$10^{-3}$ when the model parameter $\alpha$ is between 0.3 and 0.5, which correspond to partial widths of 1--50 keV. For the decays of $\Upsilon(10753) \to\eta_b \eta$ and $\eta_b\eta'$, their branching fractions depend on the $\eta$-$\eta'$ mixing angle.

Moreover, the relative ratios of the process $\Upsilon(10753)\to\eta_b\omega$ to $\Upsilon(10753)\to\Upsilon(nS)\pi^+\pi^-$ ($n=1\,,2$) are found to be in accordance with the experimental results. Our calculated results tend to support the interpretation of the $\Upsilon(10753)$ as a $4S$-$3D$ mixture. It is hoped that the present calculated results could be verified by the experiments in BelleII.


\begin{acknowledgements}\label{sec:acknowledgements}

The authors are grateful to Ziyue Bai and Yushuai Li at Lanzhou University and Qi Wu at Henan Normal University for constructive instruction in calculation programs. This work is partly supported by the National Natural Science Foundation of China under Grant Nos. 12075133, 12105153, 11835015, 12047503, and 12075288, and by the Natural Science
Foundation of Shandong Province under Grant Nos. ZR2021MA082, ZR2022ZD26, and ZR2021ME147. It is also supported by Taishan Scholar Project of Shandong Province (Grant No.tsqn202103062), the Higher Educational Youth Innovation Science and Technology Program Shandong Province (Grant No. 2020KJJ004).
	
\end{acknowledgements}
\onecolumngrid
\appendix

\section{Tensor and vector structures in Eq. \eqref{eq:msd}}\label{app:amps}

Here we only give the tensor and vector structures $N_{\mu(\nu)}$ in Eq.\eqref{eq:msd} for the Figs.~\ref{fig:feyndiags}(b)--(f). The terms $C_{\mu(\nu)}$ and $S_{\mu}$ can be readily obtained by changing the neutral $B^{(*)0}$ into corresponding charged and strange ones.

In the cases of the $\Upsilon(p)\to B^{*0}(p_1)\bar{B}^{*0}(p_2)[B^0(q)]\to \eta_b(p_4)\omega(p_3)$ depicted in the Figs.~\ref{fig:feyndiags}(b)--(f), we have 
\begin{subequations}
	\begin{align}
		N_{\mu\nu}^{S(b)} =\ii^3\int \frac{\dd[4]{q}}{(2\pi)^4}& \big[-g_{\Upsilon B^* B^*}\big(p_{1\sigma}g_{\mu\alpha}-(p_1-p_2)_\mu g_{\alpha\sigma}-p_{2\alpha}g_{\mu\sigma}\big)\big]\nonumber\\
		\times & \big[g_{B^*B^*V}(p_1+q)_\nu g_{\beta\kappa}-4f_{B^*B^*V} (p_{3\kappa}g_{\nu\beta}-p_{3\beta}g_{\nu\kappa})\big]\big[-g_{\eta_b B^*B^*}p_4^\phi (q-p_2)^\zeta \epsilon_{\zeta\tau\phi\rho}\big]\nonumber\\
		\times&  D^{\alpha\beta}(p_1\,,m_{B^{*0}})D^{\sigma\rho}(p_2\,,m_{\bar{B}^{*0}})D^{\kappa\tau}(q\,,m_{B^0})F(q\,,m_{B^{*0}})\,,\\
	N_{\mu\nu}^{D(b)} =\ii^3\int \frac{\dd[4]{q}}{(2\pi)^4}& \big[-g_{\Upsilon_1 B^* B^*}\big(p_{1\sigma}g_{\mu\alpha}-4(p_1-p_2)_\mu g_{\alpha\sigma}-p_{2\alpha}g_{\mu\sigma}\big)\big]\nonumber\\
	\times &\big[g_{B^*B^*V}(p_1+q)_\nu g_{\beta\kappa}-4f_{B^*B^*V} (p_{3\kappa}g_{\nu\beta}-p_{3\beta}g_{\nu\kappa})\big]\big[-g_{\eta_b B^*B^*}p_4^\phi (q-p_2)^\zeta \epsilon_{\zeta\tau\phi\rho}\big]\nonumber\\
	\times&  D^{\alpha\beta}(p_1\,,m_{B^{*0}})D^{\sigma\rho}(p_2\,,m_{\bar{B}^{*0}})D^{\kappa\tau}(q\,,m_{B^0})F(q\,,m_{B^{*0}})\,.
	\end{align}
\end{subequations}
\begin{subequations}
	\begin{align}
	N_{\mu\nu}^{S(c)} =\ii^3\int \frac{\dd[4]{q}}{(2\pi)^4}& \big[-g_{\Upsilon BB^*} p^\xi(p_1-p_2)^\gamma \epsilon_{\xi\mu\alpha\gamma}\big]\big[g_{B^*B^*V}(p_1+q)_\nu g_{\beta\kappa}-4f_{B^*B^*V} (p_{3\kappa}g_{\nu\beta}-p_{3\beta}g_{\nu\kappa})\big]\nonumber\\
	\times & \big[-g_{\eta_bBB^*} (q-p_2)_\tau \big] D^{\alpha\beta}(p_1\,,m_{B^{*0}})D(p_2\,,m_{\bar{B}^{*0}})D^{\kappa\tau}(q\,,m_{B^0})F(q\,,m_{B^{*0}})\,,\\
		N_{\mu\nu}^{D(c)} =\ii^3\int \frac{\dd[4]{q}}{(2\pi)^4}& \big[-g_{\Upsilon BB^*} p^\xi(p_1-p_2)^\gamma \epsilon_{\xi\mu\alpha\gamma}\big]\big[g_{B^*B^*V}(p_1+q)_\nu g_{\beta\kappa}-4f_{B^*B^*V} (p_{3\kappa}g_{\nu\beta}-p_{3\beta}g_{\nu\kappa})\big]\nonumber\\
	\times & \big[-g_{\eta_bBB^*} (q-p_2)_\tau \big] D^{\alpha\beta}(p_1\,,m_{B^{*0}})D(p_2\,,m_{\bar{B}^{*0}})D^{\kappa\tau}(q\,,m_{B^0})F(q\,,m_{B^{*0}})\,.
	\end{align}
\end{subequations}
\begin{subequations}
	\begin{align}
		N_{\mu\nu}^{S(d)} =\ii^3\int \frac{\dd[4]{q}}{(2\pi)^4}& \big[g_{\Upsilon BB^*} p^\xi(p_1-p_2)^\gamma \epsilon_{\xi\mu\sigma\gamma}\big]\big[2f_{B^*BV}p_{3}^{\lambda}(p_1+q)^\delta \epsilon_{\lambda \nu \delta \kappa}\big]\big[-g_{\eta_b B^*B^*}p_4^\phi (q-p_2)^\zeta \epsilon_{\zeta\tau\phi\rho}\big]\nonumber\\
		\times & D(p_1\,,m_{B^{*0}})D^{\sigma\rho}(p_2\,,m_{\bar{B}^{*0}})D^{\kappa\tau}(q\,,m_{B^0})F(q\,,m_{B^{*0}})\,,\\
		N_{\mu\nu}^{D(d)} =\ii^3\int \frac{\dd[4]{q}}{(2\pi)^4}& \big[g_{\Upsilon_1 BB^*} p^\xi(p_1-p_2)^\gamma \epsilon_{\xi\mu\sigma\gamma}\big]\big[2f_{B^*BV}p_{3}^{\lambda}(p_1+q)^\delta \epsilon_{\lambda \nu \delta \kappa}\big]\big[-g_{\eta_b B^*B^*}p_4^\phi (q-p_2)^\zeta \epsilon_{\zeta\tau\phi\rho}\big]\nonumber\\
		\times & D(p_1\,,m_{B^{*0}})D^{\sigma\rho}(p_2\,,m_{\bar{B}^{*0}})D^{\kappa\tau}(q\,,m_{B^0})F(q\,,m_{B^{*0}})\,.
	\end{align}
\end{subequations}
\begin{subequations}
	\begin{align}
		N_{\mu\nu}^{S(e)}=\ii^3\int \frac{\dd[4]{q}}{(2\pi)^4}& \big[-g_{\Upsilon BB} (p_1-p_2)_\mu \big]\big[2f_{B^*BV}p_{3}^{\lambda}(p_1+q)^\delta \epsilon_{\lambda \nu \delta \kappa}\big]\big[-g_{\eta_bBB^*} (q-p_2)_\tau \big]\nonumber\\
		\times & D(p_1\,,m_{B^{*0}})D(p_2\,,m_{\bar{B}^{*0}})D^{\kappa\tau}(q\,,m_{B^0})F(q\,,m_{B^{*0}})\,,\\
		N_{\mu\nu}^{D(e)}=\ii^3\int \frac{\dd[4]{q}}{(2\pi)^4}& \big[-g_{\Upsilon BB} (p_1-p_2)_\mu \big]\big[2f_{B^*BV}p_{3}^{\lambda}(p_1+q)^\delta \epsilon_{\lambda \nu \delta \kappa}\big]\big[-g_{\eta_bBB^*} (q-p_2)_\tau \big]\nonumber\\
		\times & D(p_1\,,m_{B^{*0}})D(p_2\,,m_{\bar{B}^{*0}})D^{\kappa\tau}(q\,,m_{B^0})F(q\,,m_{B^{*0}})\,.
	\end{align}
\end{subequations}
\begin{subequations}
	\begin{align}
		N_{\mu\nu}^{S(f)} =\ii^3\int \frac{\dd[4]{q}}{(2\pi)^4}& \big[g_{\Upsilon BB^*} p^\xi(p_1-p_2)^\gamma \epsilon_{\xi\mu\sigma\gamma}\big]\big[-g_{BBV}(p_1+q)_\nu\big][-g_{\eta_bBB^*}(q-p_2)_\rho] \nonumber\\
		\times & D(p_1\,,m_{B^{*0}})D^{\sigma\rho}(p_2\,,m_{\bar{B}^{*0}})D(q\,,m_{B^0})F(q\,,m_{B^0})\,,\\
		N_{\mu\nu}^{D(f)} =\ii^3\int \frac{\dd[4]{q}}{(2\pi)^4}& \big[g_{\Upsilon_1 BB^*} p^\xi(p_1-p_2)^\gamma \epsilon_{\xi\mu\sigma\gamma}\big]\big[-g_{BBV}(p_1+q)_\nu\big][-g_{\eta_bBB^*}(q-p_2)_\rho] \nonumber\\
		\times & D(p_1\,,m_{B^{*0}})D^{\sigma\rho}(p_2\,,m_{\bar{B}^{*0}})D(q\,,m_{B^0})F(q\,,m_{B^0})\,.
	\end{align}
\end{subequations}

In the cases of the $\Upsilon(p)\to B^{*0}(p_1)\bar{B}^{*0}(p_2)[B^0(q)]\to \eta_b(p_4)\omega(p_3)$ depicted in the Figs.~\ref{fig:feyndiags}(b)--(e), we have 
	\begin{subequations}
	\begin{align}
		N_{\mu}^{S(b)} =\ii^3\int \frac{\dd[4]{q}}{(2\pi)^4}& \big[-g_{\Upsilon B^* B^*}\big(p_{1\sigma}g_{\mu\alpha}-(p_1-p_2)_\mu g_{\alpha\sigma}-p_{2\alpha}g_{\mu\sigma}\big)\big]\big[-g_{B^*B^*P}p_{1}^\lambda q^\delta \epsilon_{\lambda\beta\delta\delta\kappa}\big]\nonumber\\
		\times& \big[-g_{\eta_b B^*B^*}p_4^\phi (q-p_2)^\zeta \epsilon_{\zeta\tau\phi\rho}\big] D^{\alpha\beta}(p_1\,,m_{B^{*0}})D^{\sigma\rho}(p_2\,,m_{\bar{B}^{*0}})D^{\kappa\tau}(q\,,m_{B^0})F(q\,,m_{B^{*0}})\,,\\
		N_{\mu}^{D(b)} =\ii^3\int \frac{\dd[4]{q}}{(2\pi)^4}& \big[-g_{\Upsilon_1 B^* B^*}\big(p_{1\sigma}g_{\mu\alpha}-4(p_1-p_2)_\mu g_{\alpha\sigma}-p_{2\alpha}g_{\mu\sigma}\big)\big]\big[-g_{B^*B^*P}p_{1}^\lambda q^\delta \epsilon_{\lambda\beta\delta\kappa}\big]\nonumber\\
		\times&\big[-g_{\eta_b B^*B^*}p_4^\phi (q-p_2)^\zeta \epsilon_{\zeta\tau\phi\rho}\big] D^{\alpha\beta}(p_1\,,m_{B^{*0}})D^{\sigma\rho}(p_2\,,m_{\bar{B}^{*0}})D^{\kappa\tau}(q\,,m_{B^0})F(q\,,m_{B^{*0}})\,.
	\end{align}
\end{subequations}
\begin{subequations}
	\begin{align}
		N_\mu^{S(c)} =\ii^3\int \frac{\dd[4]{q}}{(2\pi)^4}& \big[-g_{\Upsilon BB^*} p^\xi(p_1-p_2)^\gamma \epsilon_{\xi\mu\alpha\gamma}\big]\big[-g_{B^*B^*P}p_{1}^\lambda q^\delta \epsilon_{\lambda\beta\delta\kappa}\big]\big[-g_{\eta_bBB^*} (q-p_2)_\tau \big]\nonumber\\
		\times& D^{\alpha\beta}(p_1\,,m_{B^{*0}})D(p_2\,,m_{\bar{B}^{*0}})D^{\kappa\tau}(q\,,m_{B^0})F(q\,,m_{B^{*0}})\,,\\
		N_\mu^{D(c)} = \ii^3\int \frac{\dd[4]{q}}{(2\pi)^4}& \big[-g_{\Upsilon_1 BB^*} p^\xi(p_1-p_2)^\gamma \epsilon_{\xi\mu\alpha\gamma}\big]\big[-g_{B^*B^*P}p_{1}^\lambda q^\delta \epsilon_{\lambda\beta\delta\kappa}\big]\big[-g_{\eta_bBB^*} (q-p_2)_\tau \big]\nonumber\\
		\times& D^{\alpha\beta}(p_1\,,m_{B^{*0}})D(p_2\,,m_{\bar{B}^{*0}})D^{\kappa\tau}(q\,,m_{B^0})F(q\,,m_{B^{*0}})\,.
	\end{align}
\end{subequations}
\begin{subequations}
	\begin{align}
		N_\mu^{S(d)} =\ii^3\int \frac{\dd[4]{q}}{(2\pi)^4}& \big[g_{\Upsilon BB^*} p^\xi(p_1-p_2)^\gamma \epsilon_{\xi\mu\sigma\gamma}\big]\big[g_{B^*BP}p_{3\kappa}\big]\big[-g_{\eta_b B^*B^*}p_4^\phi (q-p_2)^\zeta \epsilon_{\zeta\tau\phi\rho}\big]\nonumber\\
		\times & D(p_1\,,m_{B^{*0}})D^{\sigma\rho}(p_2\,,m_{\bar{B}^{*0}})D^{\kappa\tau}(q\,,m_{B^0})F(q\,,m_{B^{*0}})\,,\\
		N_\mu^{D(d)} =\ii^3\int \frac{\dd[4]{q}}{(2\pi)^4}& \big[g_{\Upsilon_1 BB^*} p^\xi(p_1-p_2)^\gamma \epsilon_{\xi\mu\sigma\gamma}\big]\big[g_{B^*BP}p_{3\kappa}\big]\big[-g_{\eta_b B^*B^*}p_4^\phi (q-p_2)^\zeta \epsilon_{\zeta\tau\phi\rho}\big]\nonumber\\
		\times & D(p_1\,,m_{B^{*0}})D^{\sigma\rho}(p_2\,,m_{\bar{B}^{*0}})D^{\kappa\tau}(q\,,m_{B^0})F(q\,,m_{B^{*0}})\,.
	\end{align}
\end{subequations}
\begin{subequations}
	\begin{align}
		N_\mu^{S(e)}=\ii^3\int \frac{\dd[4]{q}}{(2\pi)^4}& \big[-g_{\Upsilon BB} (p_1-p_2)_\mu \big]\big[g_{B^*BP}p_{3\kappa}\big]\big[-g_{\eta_bBB^*} (q-p_2)_\tau \big]\nonumber\\
		\times & D(p_1\,,m_{B^{*0}})D(p_2\,,m_{\bar{B}^{*0}})D^{\kappa\tau}(q\,,m_{B^0})F(q\,,m_{B^{*0}})\,,\\
		N_\mu^{D(e)}=\ii^3\int \frac{\dd[4]{q}}{(2\pi)^4}& \big[-g_{\Upsilon_1 BB} (p_1-p_2)_\mu \big]\big[g_{B^*BP}p_{3\kappa}\big]\big[-g_{\eta_bBB^*} (q-p_2)_\tau \big]\nonumber\\
		\times & D(p_1\,,m_{B^{*0}})D(p_2\,,m_{\bar{B}^{*0}})D^{\kappa\tau}(q\,,m_{B^0})F(q\,,m_{B^{*0}})\,.
	\end{align}
\end{subequations}

\twocolumngrid	
\bibliography{particlePhys.bib}
\end{document}